\documentclass[12pt]{article}

\newcommand{\JMPcite}[1]{\kern-.2em${}^{\hbox{\scriptsize\cite{#1}}}$}%

\newcommand{\be}{\begin{equation}}
\newcommand{\ee}{\end{equation}}
\newcommand{\bea}{\begin{eqnarray}}
\newcommand{\eea}{\end{eqnarray}}

\newcommand \jj {{\underline{j}}}

\newcommand\proof{\par\noindent{\it Proof}\/:\ }

\newcommand{\tr} {{\rm tr}}

\newcommand{\dist} {{\rm dist}}

\def\zed{{\mathbf{Z}}}
\def\C{{\mathbf{C}}}
\def\reff#1{(\ref{#1})}
\def\bbbc{{\mathchoice {\setbox0=\hbox{$\displaystyle\rm C$}\hbox{\hbox
to0pt{\kern0.4\wd0\vrule height0.9\ht0\hss}\box0}}
{\setbox0=\hbox{$\textstyle\rm C$}\hbox{\hbox
to0pt{\kern0.4\wd0\vrule height0.9\ht0\hss}\box0}}
{\setbox0=\hbox{$\scriptstyle\rm C$}\hbox{\hbox
to0pt{\kern0.4\wd0\vrule height0.9\ht0\hss}\box0}}
{\setbox0=\hbox{$\scriptscriptstyle\rm C$}\hbox{\hbox
to0pt{\kern0.4\wd0\vrule height0.9\ht0\hss}\box0}}}}
\newtheorem{proposition}{Proposition}
\newtheorem{lemma}{Lemma}
\newtheorem{theorem}{Theorem}

\def\bbbr{{\rm I\!R}} 

\def\bbbe{{\rm I\!E}}

\def\bbbc{{\rm I\!C}}

\begin{document}
\title{Data compression limit for an information source of interacting qubits}
\author{Nilanjana Datta and Yuri Suhov\\
{Statistical Laboratory}\\
DPMMS, Centre for Mathematical Sciences,\\
University of Cambridge\\
Cambridge CB3 OWB.}
\date{July 11, 2002}
\maketitle

\bigskip

\bigskip

\begin{abstract}
A system of interacting qubits can be viewed as a non-i.i.d quantum
information source. A possible model of such a source is provided by
a quantum spin system, in which spin-1/2 particles located at sites of a 
lattice interact with each other. We establish the limit
for the compression of information from such a source and show that
asymptotically it is given by the von Neumann entropy rate. Our result
can be viewed as a quantum analogue of Shannon's noiseless coding theorem
for a class of {\em{non - i.i.d.}} quantum information sources.
\end{abstract}

\newpage

\section{Introduction} 

In this paper we study the issue of 
compression of information from a particular class of quantum
information sources, formed by systems of interacting qubits [see Section
\ref{setup} for details]. Our aim is to quantify 
the minimal physical resources necessary to store the output 
from such a source or to
transmit it through a noiseless channel. 
We shall use the words message, signal 
and output from a source interchangeably. 
The parameter that we 
minimise is the dimension of the Hilbert space to which a typical signal
can be projected (i.e., ``compressed'') with high fidelity. 
In addition, it is expected that the interaction between qubits in the systems
under consideration yields highly-entangled states; this is a motivation
for the present work, even though the issue of entanglement is not
discussed here.

The analysis that follows shows that the data
compression limit for output from such a source 
is given by the von Neumann entropy rate. This result can be 
viewed as a {\em{quantum analogue}}
of {{Shannon's noiseless coding theorem}} \JMPcite{shannon1} for our class 
of {\em{non-i.i.d}} quantum sources. It can be considered
as an extension of Schumacher's coding theorem \JMPcite{schumacher}.

Shannon's noiseless coding theorem {{quantifies}} the extent to
which one can {{compress}} the information being produced by a
classical information source.
A standard model of such a source is described by a sequence of 
random variables 
$X_1, X_2,\ldots, X_n$ whose values $x_1,x_2,\ldots,x_n$ represent the
output of the source. For simplicity, consider random
variables which take values from a finite alphabet of symbols or letters
(extensions to infinite alphabets also hold). Let 
${\underline{X}} := (X_1,X_2,\ldots X_n)$ denote the sequence
of random variables representing the source and
${\underline{x}} := (x_1,x_2,\ldots x_n)$  the values that it takes.
The source is described by a set of probabilities 
$$
p({\underline{x}}) := {\hbox{Prob}}\,({\underline{X}}={\underline{x}}).
$$
An {{ i.i.d classical source}} is one for which 
the random variables $X_1, X_2 \ldots X_n$ are independent and identically
distributed. In this case
$$
p({\underline{x}}) := p(x_1)p(x_2)\ldots p(x_n),
$$
where $\{p(x)\}$ is the single symbol distribution.

The main ingredient of Shannon's noiseless coding theorem is the Shannon
entropy given by
$$
H({\underline{X}}):= - \sum_{{\underline{x}}} p({\underline{x}}) \log_2\,
 p({\underline{x}});
$$
for an i.i.d. source this reduces to $H({\underline{X}}) = n H(X)$, where
$$
H({{X}}):= - \sum_{{{x}}} p({{x}}) \log_2\,
 p({x}).
$$
In classical information theory one {{encodes}} the signal
from a source into a string of binary digits (or bits). 
For purposes
of storage and transmission, the aim is to encode  
the messages with sequences that 
are as short as possible. 

An information source (classical or quantum) has redundancy, 
in the sense that certain
outputs occur more frequently than the rest. This fact can be used 
to compress the source output: data compression is
achieved by assigning shorter descriptions to the most frequent outputs
of the source. The compression of data from a classical information
source works as follows \JMPcite{cover, nielson}: A compression map, $C^n$, 
of rate $R$ takes a sequence ${\underline{x}} = (x_1, \ldots, x_n)$
of length $n$ to a binary string of length $[nR]$ (the symbol 
$[\cdot]$ denoting the integer part). A decompression
map, $D^n$, takes a binary string of length $[nR]$ to a 
string of symbols of length $n$. 
The compression scheme is said to be reliable if with probability
approaching one, as $n \rightarrow \infty$, 
$D^n\left(C^n({\underline x})\right) 
= {\underline x}.
$

Shannon's noiseless coding theorem indicates how well such a 
compression scheme works.
More precisely, it asserts that for a large class of
sources (i.e., stationary and ergodic), 
the mean length of encoded bit sequences is asymptotically given by the 
Shannon entropy, $H({\underline{X}})$. More precisely, the data compression
limit, which is the limiting number of bits per symbol, is given 
by the Shannon entropy rate:
$$
h:= {\lim_{n \rightarrow \infty}} \,\frac{1}{n}\,
H({{X_1\ldots X_n}}). 
$$
An attempt to represent the 
source using fewer bits
than this would result in a high probability of error when the
information is decompressed. Hence a compression scheme of rate $R$
is reliable only if $R > h$.

        A {\em{quantum information source}} is defined in this paper 
by a set of distinguishable quantum-mechanical states $|\psi_j\rangle$, 
i.e., orthonormal
vectors from a given Hilbert space,
and a set of corresponding probabilities $\{\kappa_j\}$. We interpret the 
$|\psi_j\rangle$'s as signals of the source and the
$\kappa_j$'s as the probabilities with which the signals are produced.
Such a definition arises naturally from the density matrix formalism 
where a quantum-mechanical system is described by a convex linear
combination of pure states:
$$\rho = \sum_j \kappa_j |\psi_j\rangle \langle\psi_j|.
$$
Here the $|\psi_j\rangle$'s are identified as the 
orthonormal eigenvectors of $\rho$.
The eigenvalue of 
$\rho$ corresponding to $|\psi_j\rangle$ is $\kappa_j$, and we
have
$$ \kappa_j \ge 0 \quad {\hbox{and}} \quad \sum_j \kappa_j =1. 
\footnote{More generally, one can consider a decomposition of $\rho$ in terms
of non-orthogonal states:
$$\rho = \sum_i p_i |\phi_i\rangle \langle\phi_i|,$$
where $p_i \ge 0$, $\sum_i p_i =1$. Such a decomposition is non-unique.
Moreover, it is known that non-orthogonal
states cannot be reliably distinguished. Hence, in the sequel, we focus
on orthogonal decompositions. An exception is Theorem \ref{nonortho},
which gives the limiting fidelity of the data compression scheme
for general non-orthogonal decompositions .}$$

More precisely, we deal with a sequence of $2^n \times 2^n$ 
density matrices $\rho_n$, $n \rightarrow \infty$,
and relate asymptotic properties of their
eigenvalues $\kappa_j^{(n)}$ to the von Neumann entropy rate.
The {\em{von Neumann entropy}} of a density matrix $\rho_n$ is given by
\bea
S(\rho_n) = - \tr\, \rho_n\, \log_2\, \rho_n = - \sum_j \,\kappa_j^{(n)}\, 
\log_2 \, \kappa_j^{(n)},
\label{vnentropy}
\eea
and the von Neumann entropy rate by 
$$
h = \lim_{n \rightarrow \infty} \frac{1}{n} \, S(\rho_n).
$$

A useful example is an i.i.d. case where $\rho_n$ acts on a 
tensor product Hilbert space ${\cal H}_n = {\cal K}^{\otimes n}$
and is given by
$$
\rho_n = \pi^{\otimes n}.
$$
Here ${\cal K}$ is a fixed Hilbert
space (representing an ``elementary'' quantum subsystem) and 
$\pi$ is a density matrix acting on ${\cal K}$:
$$
\pi = \sum_i q_i |\phi_i\rangle \langle \phi_i |.$$ The
eigenvectors $|\psi_j^{(n)}\rangle$ of $\rho_n$ are tensor products
$$
|\psi_j^{(n)}\rangle= |\phi_{j_1} \rangle \otimes |\phi_{j_2} \rangle 
\ldots \otimes 
|\phi_{j_n}\rangle.$$
and its eigenvalues $\kappa_j^{(n)}$ are given by 
$$
\kappa_j^{(n)} =q_{j_1} \ldots q_{j_n}.
$$
This provides a convenient identification of label $j$ as a 
``classical string'' $(j_1, \ldots, j_n)$ which will
be emphasized by the notation ${\underline j}$ below. 
The von Neumann entropy is in this case $S( \pi^{\otimes n}) = n S(\pi).$

In the case where $\dim\,{\cal K}  = 2$, the space ${\cal H}_n$
represents a system of $n$ qubits. In analogy with classical
data compression, it is desirable to represent typical outputs, 
$|\psi_j^{(n)}\rangle$, by vectors from a lower dimensional Hilbert space, 
thereby reducing the number of qubits needed for the source description.

In his seminal paper \JMPcite{schumacher}, Schumacher proved that the number of
qubits necessary to represent, reliably, the signal from an 
i.i.d quantum information source is asymptotically given
by the von Neumann entropy. More precisely, 
there exists
a reliable compression scheme of rate $R$ only when $R > S(\pi)$ 
(under a suitable definition of fidelity). Schumacher's approach was 
developed further in \JMPcite{jozsa, mixed}.
Extensions of Schumacher's theorem to some classes of quantum sources 
with memory have been established by Petz et al (see \JMPcite{petz,
pm} and references therein).

As was said before, in this paper we consider data compression for a class of 
quantum information
sources which are modelled by a system of interacting quantum spins.
This is an example of a quantum system with a strong coupling between
the spins and with the environment and it does not fall into the classes
of sources considered in the literature before. Besides, we consider
properties of eigenvalues $\kappa_j^{(n)}$ which hold asymptotically
with probability one; this is a refinement of results obtained in 
\JMPcite{petz, pm}. From the probabilistic point of view, our result is an
analogue of the Shannon-McMillan-Breiman theorem (which is a version
of the Law of large numbers), see \JMPcite{cover}.

Even though we consider so--called quantum spin systems as models of a
quantum source in this paper, our results also hold for sources
modelled by quantum lattice gases, where the statistics 
(Bose or Fermi) of the particles is taken into account.

Models of quantum information sources, based on large systems of
interacting spins or particles, are being used increasingly
in experiments with entanglement \JMPcite{expts1, expts2}, 
as well as in theoretical research \JMPcite{korepin, theory}.
As mentioned
before, our main result can be viewed as an extension of Schumacher's coding 
theorem to this class of sources. Section \ref{setup}
contains a mathematical description of the class of 
systems under consideration. In Section \ref{results} we 
prove that the data compression limit for 
such a class is given by the von
Neumann entropy rate [see \reff{vnrate}]. The proof of the main 
theorem, which yields the data compression limit, 
is given in Section \ref{proofs}.

\section{Quantum spin systems}
\label{setup}

We consider a quantum-mechanical system on a $d$-dimensional lattice 
$\zed^{d}$, with a spin-$1/2$ particle attached to each site 
of the lattice. The particle can be either in an up-spin state (denoted
by $|\uparrow\rangle$ ) or a down-spin state (denoted
by $|\downarrow\rangle$ ). Hence, to each lattice site $x \in \zed^d$
is associated a Hilbert space ${\cal{K}}_x$ which is isomorphic to
${\cal{K}}=\C^2$, the single-qubit Hilbert space. For any finite subset $X \subset \zed^d$, the corresponding Hilbert 
space is given by
$$      
{\cal H}_{X}=\otimes_{x \in X}{\cal K}_{x} = (\C^2)^{\otimes |X|}.
$$
Here, and below, $|B|$
stands for the number of elements in a finite set $B$. Furthermore, we 
denote by ${\cal A}_X$ the algebra of $2^{|X|} \times 
2^{|X|}$ matrices acting in  ${\cal H}_X$  -- 
the local observable algebra. To each site $x$ of the lattice, we
associate a variable $j_x \in \{1,-1\}$ such that
$j_x = 1 (-1)$ when the spin at $x$ is $\uparrow (\downarrow)$.
A configuration $\omega_\Lambda$ in a finite volume 
$\Lambda \subset \zed^d$ is an assignment
 $\{j_x, x \in \Lambda\}$ of $j_x$ to each $x \in \Lambda$;
the set of configurations $\{\omega_\Lambda\}$ provides labels for a
quasiclassical basis $\{|\omega_\Lambda\rangle\}$ in
${\cal H}_{\Lambda} = (\C^2)^{\otimes |\Lambda|}.$

The physics of the system is described by an  
interaction, $\Phi = \{\Phi_X\}$,
which is a map taking finite subsets $X\subset\zed^{d}$ to (self-adjoint) 
operators $\Phi_X$ from ${\cal A}_X$; see \JMPcite{DFFR}. 
We study quantum systems that are small perturbations of 
classical ones. That is, we consider interactions of the form
$\Phi = \Phi_0 + Q$, with 
\begin{equation}
\Phi_{X} = \Phi_{0X} +  Q_X,
\end{equation}
where, for all $X$, $\Phi_{0X}$ is diagonal in the quasiclassical
basis $\{| \omega_\Lambda \rangle \}$ and $Q_X$ is small in norm 
(see below). We will write $\Phi_0 = \{\Phi_{0X}\}$ and $Q =  \{Q_X\}$.

The corresponding Hamiltonian 
$H_{\Lambda}= \sum_{X \subset \Lambda} \Phi_{X}$ of a system 
confined to a finite volume $\Lambda \subset \zed^{d}$ 
is written as a sum
\begin{equation}
H_{\Lambda}=H_{0\Lambda} +  V_{\Lambda}, \label{ham}
\end{equation}
where $H_{0\Lambda} := \sum_{X \subset \Lambda} \Phi_{0X}$, 
$V_{\Lambda} 
:= \sum_{X \subset \Lambda} Q_X$. 
We make the following assumptions:

{\em{(i)}} We consider translation-invariant interactions (for details,
see \JMPcite{DFF}) i.e., 
$\Phi_{X + a} \, \simeq \,\Phi_X$, for all  finite $X\subset
{\zed^{d}}$ and $a\in {\zed^{d}}$. The {\em range}\/ of the interaction is
defined as the supremum of the 
diameters of sets $X$ from 
$\{X\subset\zed^d : X \ni 0 \hbox{ and } \Phi_X \neq 0\}$.   
We use the $\ell^\infty$-diameter 
\begin{equation} 
{\hbox{diam M}} \;:=\; \max_{x,y \in M} \max_{1\leq i \leq d} 
|x_i - y_i|\;,\label{diam}  
\end{equation} 
and consider $\Phi$ to be of a finite range, i.e., 
with $R < \infty$.
\smallskip

\noindent
{\em{(ii)}} The classical part $\Phi_{0X}$ of $\Phi_{X}$ 
can be considered as a real-valued function on the set of configurations
$\omega_X$ in $X$ ( i.e., an assignment $\{j_x, x \in X\}$).
It is convenient to think of $\Phi_{0X}$ as a function of the 
infinite-volume configuration $w \equiv \{j_x, x \in \zed^d\}$, which
depends on its restriction $w_X$ only. Similarly, 
$H_{0\Lambda}$ is a real-valued function of $\omega$ depending on
$\omega_\Lambda$ only. We call an infinite-volume configuration
$\omega \equiv \{j_x, x \in \zed^d\}$ {\em{periodic}} 
if $j_x = j_{x + a^{(i)}}$,
$i=1,\ldots,d$, for all $x \in \zed^d$ and a given collection of periods 
$a^{(i)}= n^{(i)}\,e^{(i)}$ where $e^{(i)} = (0,\ldots,1,\ldots,0)\in \zed^d$
(entry $1$ at position $i$) and $n^{(i)}$ is a given integer. A periodic
$\omega$ is called a ground state configuration for $\Phi_0$ if 
$$
\liminf_{\Lambda \nearrow \zed^d}
\frac{1}{|\Lambda|} \left[ H_{0\Lambda} (\omega^\prime) 
-  H_{0\Lambda} (\omega)\right] \ge 0\,\footnote{
Here and below, the symbol $\Lambda \nearrow \zed^d$ is used for the
{\em{thermodynamical limit}}, taken along a sequence of growing finite
volumes $\Lambda_1 \subset \Lambda_2 \subset \ldots \subset \zed^d$
of ``nice'' shape (e.g., hypercubes $\Lambda_n = [-n,n]^d \cap \zed^d$),
with $\cup_n \Lambda_n = \zed^d$.},$$
for any infinite-volume configuration $\omega^\prime$.

We assume that $\Phi_0$ has a {\it finite} number of
periodic classical ground states, $\sigma^{(1)}, \ldots, \sigma^{(m)}$,
and satisfies the so-called Peierls condition \JMPcite{holsla78}. 
The latter is a condition for stability of the 
ground states relative to ``local'' perturbations. (See \JMPcite{DFF,bku} 
and references therein for details.)
\smallskip

\noindent
{\em{(iii)}} The term $V_\Lambda$ is a quantum perturbation 
$\left({\rm i.e.,}\,[H_{0\Lambda}, V_\Lambda]\ne 0 \right)$, with
$$
||Q_X|| \le c \lambda^{s(X)}
$$
for some constant $c$ and some $0<\lambda<1$. Here $s(X)$ denotes the number
of sites in the smallest connected subset of the lattice containing $X$.
We consider $\lambda$ as the perturbation
parameter.

Assumptions {\em{(i) - (iii)}} constitute the framework of the 
so-called quantum Pirogov-Sinai theory \cite{bku, DFF, pir78, 
pirsin75, pirsin76}. 
\medskip
 
We fix a boundary condition outside volume $\Lambda$, i.e., 
assume that the configuration on $\Lambda^c
:=
\zed^d \setminus \Lambda$ coincides with a fixed reference
configuration $\sigma$, which is one of the 
periodic ground states $\sigma^{(1)}, \ldots, \sigma^{(m)}$ 
of $H_{0\Lambda}$. 

Since the interaction is of a finite range, the spins in 
$\Lambda$ interact only with those spins in ${\Lambda^c}$
that are in the envelopping volume $\Lambda^\partial$:
$$
\Lambda^\partial = \{ i \in {\Lambda^c}\,:\,\, \dist(i,j) \le R \,
\,{\hbox{for some}}\,
j \in \Lambda\}.
$$
Let $P^{\sigma}_\Lambda$ be 
the orthogonal projection onto the subspace  ${\cal H}^\sigma_\Lambda \subset 
{\cal{H}}_{\Lambda\cup\Lambda^\partial}$ of dimension $2^{|\Lambda|}$, 
spanned by states for which the configuration on $\Lambda^\partial$ is fixed
to $\sigma$.  Then the Hamiltonian governing the spin system in 
$\Lambda$ under the boundary condition
$\sigma$ is given by
$$H_\Lambda^\sigma = P^{\sigma}_\Lambda\,H_{\Lambda\cup\Lambda^\partial}
\,P^{\sigma}_\Lambda \equiv \sum_{X:\atop{X \cap \Lambda \ne \emptyset}}
 P^{\sigma}_\Lambda\,\Phi_X\,P^{\sigma}_\Lambda .
$$
The spin system with Hamiltonian $H_\Lambda^\sigma$
can be viewed as a system of interacting spins entangled with
its environment. It is considered at
a finite but low temperature. Due to the interaction between spins,
the density matrix cannot be written as a tensor product of the density
matrices of the individual spins and hence the quantum information source 
is {{non-i.i.d}}. The density matrix is written in the standard
Gibbsian form:
\be
\rho^{\sigma, \Lambda} = \frac{e^{-\beta H_\Lambda^\sigma}}
{\Xi^{\sigma, \Lambda}},\label{rhogibbs}
\ee
where $\beta >0$ is the inverse temperature. 
The denominator on the RHS of \reff{rhogibbs} is the partition function:
$$
\Xi^{\sigma, \Lambda} = {\tr_{{\cal H}^\sigma_\Lambda}\,e^{-\beta H_\Lambda^\sigma}}.
$$
The expectation of an observable 
${\bf{A}} \in {\cal A}_{\Lambda\cup \Lambda'}$ in
the Gibbs state $\rho^{\sigma, \Lambda}$ is given by 
\begin{equation} 
\langle {\bf{A}} \rangle^\sigma_\Lambda 
\equiv \tr_{{\cal H}^\sigma_\Lambda}\,
\rho^{\sigma, \Lambda}\,{\bf{A}}
 =\frac{1}{\Xi^{\sigma, \Lambda}}\tr_{{\cal H}^\sigma_\Lambda} 
{\bf{A}}\, \exp\left({-\beta H_\Lambda^\sigma}\right). 
\label{fs.60} 
\end{equation}
Here and below, the trace is taken in the space ${\cal H}^\sigma_\Lambda$;
for notational simplicity, the subscript ${\cal H}^\sigma_\Lambda$ will
often be omitted.
For ${\bf{A}} := \exp(i \tau H_\Lambda)$, where $\tau \in \bbbr$, \reff{fs.60}
yields the characteristic function for the eigenvalues of the
Hamiltonian $H_\Lambda^\sigma$:
\be
\varphi^{\sigma, \Lambda} (\tau) := 
\langle e^{i \tau H_\Lambda^\sigma}\rangle^\sigma_\Lambda.
\label{varphi}
\ee
The eigenvalues $\kappa_j^{\sigma, \Lambda}$ of $\rho^{\sigma, \Lambda}$
can be written as
\be
\kappa_j^{\sigma, \Lambda} = \frac{1}{\Xi^{\sigma, \Lambda}}
\, \langle \psi_j^{\sigma, \Lambda} | e^{-\beta H_\Lambda^\sigma}| 
\psi_j^{\sigma, \Lambda} \rangle= \frac{1}{\Xi^{\sigma, \Lambda}}
\, \exp \,\left(-\beta \langle \psi_j^{\sigma, \Lambda}|  H_\Lambda^\sigma|
\psi_j^{\sigma, \Lambda} \rangle \right),
\label{seven'}
\ee
where $|\psi_1^{\sigma, \Lambda}\rangle, \ldots, 
|\psi_{2^{|\Lambda|}}^{\sigma, \Lambda}\rangle$
are the orthonormal eigenvectors of $\rho^{\sigma,\Lambda}$ (sometimes denoted by
$\psi_1^{\sigma, \Lambda}$, $\ldots,\psi_{2^{|\Lambda|}}^{\sigma, \Lambda}$).
The eigenvalues $\kappa_j^{\sigma, \Lambda}$ satisfy
\be
 \sum_j \kappa_j^{\sigma, \Lambda} =1.
\label{sumkappa}
\ee
The von Neumann entropy of $\rho^{\sigma, \Lambda}$ 
is given by 
\bea
S(\rho^{\sigma, \Lambda}) &=& - \tr \rho^{\sigma, \Lambda}\, \log_2\, 
\rho^{\sigma, \Lambda} \nonumber\\
        &=& - \sum_j \kappa_j^{\sigma, \Lambda}\, \log_2 \,
\kappa_j^{\sigma, \Lambda}.
\label{vnentropy2}
\eea
The  von Neumann entropy rate in this case is defined as
\bea
h &=& \lim_{\Lambda \nearrow\zed^d} \frac{S(\rho^{\sigma, \Lambda})}{|\Lambda|}
\nonumber\\
&=& c_0 \,\lim_{\Lambda \nearrow\zed^d} \tr\, \rho^{\sigma, \Lambda}\,
\left(\beta\frac{H^\sigma_\Lambda}{|\Lambda|} + \frac{1}{|\Lambda|} 
\, \log_e \Xi^{\sigma, \Lambda}\right)\nonumber\\
&=& \beta c_0 \, (g^{(\sigma)} - f)
\label{vnrate}
\eea
where $c_0 = \log_2 e$ and $f$ and $g^{(\sigma)}$ are standard
thermodynamical functions (the free energy and the infinite volume
energy per lattice site):
\bea
f &=& \lim_{\Lambda \nearrow\zed^d} \frac{-1}{\beta |\Lambda|} 
\log_e \Xi^{\sigma,\Lambda}; \label{free}\\
g^{(\sigma)} &=&  \lim_{\Lambda \nearrow\zed^d} \langle 
\frac{H_\Lambda^\sigma}{|\Lambda|}\rangle^\sigma_\Lambda.
\label{gsig}
\eea
We see that the von Neumann entropy rate $h$ is well-defined if 
the above limits, \reff{free} and \reff{gsig}, exist.
The following theorem, proved in \JMPcite{DFF}, states that these
limits do exist for the class of quantum spin systems 
under consideration.
\begin{proposition} 
\label{psresults}
Under the above assumptions, for $\beta$ large and $\lambda$ small
enough, the limits \reff{free} and \reff{gsig} exist.
\end{proposition}
{\em{Remark}}: In this paper we deal with a sequence of density
matrices $\rho^{\sigma, \Lambda}$, ${\Lambda \nearrow\zed^d}$,
not generated by a single state of a quasi-local algebra (see
e.g. \JMPcite{bratteli}). This puts us in a context different from
that considered e.g. in \JMPcite{lieb}. Hence we need 
Proposition \ref{psresults} to guarantee the existence of the
von Neumann entropy rate.

In view of \reff{sumkappa}, the eigenvalues $\kappa_j^{\sigma, \Lambda}$, 
$1\le j\le 2^{|\Lambda|}$, 
can be interpreted as the 
probabilities of the system being in the states 
$|\psi_j^{\sigma, \Lambda} \rangle$.
Let ${\cal P}^{\sigma, \Lambda}$ be the corresponding 
probability distribution
and consider a random variable $K^{\sigma, \Lambda}$ which 
takes a value 
$\kappa_j^{\sigma, \Lambda}$ with probability $\kappa_j^{\sigma, \Lambda}$:
$$
K^{\sigma, \Lambda}(\psi_j^{\sigma, \Lambda}) = \kappa_j^{\sigma, \Lambda}
\quad ; \quad 
{\cal P}^{\sigma, \Lambda}(K^{\sigma, \Lambda}
=\kappa_j^{\sigma, \Lambda})
= \kappa_j^{\sigma, \Lambda}.
$$
The data compression limit is related to asymptotical properties
of random variables $K^{\sigma, \Lambda}$ as 
$\Lambda \nearrow\zed^d$.


\section{Data Compression Limit}
\label{results}

The main result of the paper is the following theorem.

\begin{theorem}
\label{mainresult}
Under the above assumptions, for 
 $\beta$ large and $\lambda$ small enough, for all $\delta > 0$
\be
\lim_{\Lambda \nearrow\zed^d} {{\cal P}}^{\sigma, \Lambda} 
\left( |\frac{-1}{|\Lambda|} \log_2 K^{\sigma, \Lambda} -h |
\le \delta \right) =
\lim_{\Lambda \nearrow\zed^d} \sum_j \kappa_j^{\sigma, \Lambda}
\, {\mathbf{1}}\left(| \frac{- 1}{|\Lambda|} \, \log_2 \kappa_j^{\sigma, \Lambda}
 - h | \le \delta \right)
= 1,\label{one5}
\ee
where ${\mathbf{1}}(\cdot)$ denotes an indicator function.
\end{theorem}

Note that 
\be
{\bbbe}_{{\cal P}}^{\sigma, \Lambda}  \left(-  \frac{1}{|\Lambda|} 
\log_2 K^{\sigma, \Lambda} \right) = \sum_j \kappa_j^{\sigma, \Lambda} \left(
-  \frac{1}{ |\Lambda|} \log_2 \kappa_j^{\sigma, \Lambda}\right)
= \frac{S(\rho^{\sigma,\Lambda})}{|\Lambda|},
\ee
where ${\bbbe}_{{\cal P}}^{\sigma, \Lambda}(\cdot)$ denotes the 
expectation value with respect to the probability distribution
${\cal P}^{\sigma, \Lambda}$.
Hence, 
$$
\lim_{\Lambda \nearrow\zed^d} {{\bbbe}}_{{\cal P}}^{\sigma, \Lambda} = h,
$$
and Theorem \ref{mainresult} gives a Law of large numbers
for random variables $ \,(- \log_2\,K^{\sigma, \Lambda})$. 

The proof of Theorem \ref{mainresult} is given in 
Section \ref{proofs}. Here we discuss some of its 
consequences.
The statement of the theorem can be alternatively expressed as follows:
$\forall$ $\delta >0$\smallskip
\be
\lim_{\Lambda \nearrow\zed^d}{{\cal P}}^{\sigma, \Lambda}
\left( 2^{-|\Lambda| (h + \delta)} \le K^{\sigma, \Lambda} \le
2^{-|\Lambda| (h - \delta)} \right) = 1.
\label{eps}
\ee
In other words, $\forall$ $\epsilon >0$ and for $\Lambda$ large enough, 
the eigenvalues $\kappa_j^{\sigma, \Lambda}$ of $\rho^{\sigma, \Lambda}$
satisfy 
\be
 2^{-|\Lambda| (h + \delta)} \le \kappa_j^{\sigma, \Lambda} \le
2^{-|\Lambda| (h - \delta)}
\label{kbound}
\ee
with probability $ \ge (1-\epsilon)$. That is, the eigenstates
$|\psi_j^{\sigma, \Lambda}\rangle$ that correspond to eigenvalues  
$\kappa_j^{\sigma, \Lambda}$ satisfying \reff{kbound} are those which
occur most frequently. We refer to them as typical states
(or more precisely, $\delta$-typical states). Let 
${\cal M}_\delta^{\sigma, \Lambda}$ be the subspace spanned
by such states:
\be
{\cal M}_\delta^{\sigma,\Lambda} := {\rm span}\,\{|\psi_j^{\sigma, \Lambda}
\rangle\, : \,\reff{kbound}
\,\, {\hbox{holds}}\}
\label{mdelta}
\ee
and $|{\cal M}_\delta^{\sigma, \Lambda}|$ denote the dimension of this subspace. 
The following lemma establishes the growth rate of 
$|{\cal M}_\delta^{\sigma, \Lambda}|$.
\begin{lemma}
\label{asympt}
For all $\delta >0$
\be
\lim_{\Lambda \nearrow\zed^d}\frac{1}{|\Lambda|}\, \log_2\,
|{\cal M}_\delta^{\sigma,\Lambda}|
=h.
\label{two0}
\ee
\end{lemma}

\proof  We follow a standard information-theoretical 
argument (see e.g. \JMPcite{nielson}).
From \reff{eps} it follows that the probability of a state being 
$\delta$-typical is at least $(1-\epsilon)$ in the limit
$\Lambda \nearrow \zed^d$:
\be
\liminf \, \sum_j^{(\delta)}  \kappa_j^{\sigma, \Lambda} 
\ge 1-\epsilon. 
\label{lbd}
\ee
where the sum $\sum\limits_j^{(\delta)}$ is over those $j$'s for 
which $\kappa_j^{\sigma, \Lambda} $
satisfies \reff{kbound}, i.e., $| \psi_j^{\sigma, \Lambda}\rangle
\in {\cal M}_\delta^{\sigma,\Lambda}$. From \reff{lbd} (and the definition \reff{mdelta}
of the set ${\cal M}_\delta^{\sigma,\Lambda}$) 
we deduce that $\forall$ $\epsilon >0$,
\be
 1-\epsilon\le \sum_j^{(\delta)}  \kappa_j^{\sigma, \Lambda} 
\le 2^{-|\Lambda| (h - \delta)}\, | {\cal M}_\delta^{\sigma,\Lambda}|.
\label{bd2}
\ee
Also, from \reff{sumkappa} and \reff{mdelta} we have
\be
 2^{-|\Lambda| (h + \delta)}\, | {\cal M}_\delta^{\sigma,\Lambda}| \le \sum_j^{(\delta)}
\kappa_j^{\sigma, \Lambda} \le 1.
\label{bd3}
\ee
From \reff{bd2} and \reff{bd3} it follows that
$$
2^{|\Lambda| (h - \delta)} \le | {\cal M}_\delta^{\sigma,\Lambda}|
\le 2^{|\Lambda| (h + \delta)}.
$$
Since this holds for all $\epsilon > 0$, we conclude that
\be
\lim \, \sup \, \frac{1}{|\Lambda|} \, \log_2 \,
| {\cal M}_\delta^{\sigma,\Lambda}| \le h + \delta,\quad \quad
\lim \, \inf  \, \frac{1}{|\Lambda|} \, \log_2 \,
| {\cal M}_\delta^{\sigma,\Lambda}| \ge h - \delta.
\ee
Moreover, since $\delta$ is arbitrary, 
$$
\lim \, \sup  \, \frac{1}{|\Lambda|} \, \log_2 \, 
| {\cal M}_\delta^{\sigma,\Lambda}|=\lim \, \inf  \, \frac{1}{|\Lambda|} 
\, \log_2 \, | {\cal M}_\delta^{\sigma,\Lambda}|.
$$
Hence $\lim_{\Lambda \nearrow\zed^d} \, \frac{1}{|\Lambda|} 
\, \log_2 \, | 
{\cal M}_\delta^{\sigma,\Lambda}|$ exists and
is given by \reff{two0}.

\begin{lemma}
\label{compress}
Consider a quantum information source described by the density matrix
$\rho^{\sigma, \Lambda}$:
$$
\rho^{\sigma, \Lambda}\equiv 
\frac{e^{-\beta H_\Lambda^\sigma}}{\Xi^{\sigma,\Lambda}}
= \sum_j   \kappa_j^{\sigma, \Lambda} \,|\psi_j^{\sigma, \Lambda} \rangle
\langle\psi_j^{\sigma, \Lambda} |.
$$
Let $h$ be the von Neumann entropy rate [see \reff{vnrate}]. 
If $R>h$ then there exists
a reliable compression scheme of rate $R$.
\end{lemma}

\proof
Since there are {{at most}} $2^{|\Lambda| h}$ $\delta$-typical states
(see Lemma \ref{asympt}), one requires at most ${[|\Lambda| h]}$ qubits 
to uniquely identify a 
$\delta$-typical state. The data can be compressed as follows:

Map each $\delta$-typical state $|\psi_j^{\sigma, \Lambda}\rangle $
to a quasiclassical state 
$ 
|{\underline x}\rangle = |x_1\rangle\otimes | 
x_2\rangle\otimes\ldots\otimes |x_{[|\Lambda| h]}\rangle,
$
where ${\underline x}$ is a binary string of length $[|\Lambda|h]$:
$${\underline x} = (x_1, x_2, \ldots , x_{[|\Lambda| h]}) \in 
\{0,1\}^{[|\Lambda| h]}.$$

Clearly, this can be done in a one-to-one fashion, enabling us to
recover any $\delta$-typical state. In other words, 
the information contained in $|\Lambda|$ interacting qubits
is compressed into ${[|\Lambda| h]}$ non-interacting qubits, which
can be later decompressed unambiguously.
In the limit ${\Lambda \nearrow\zed^d}$ this scheme succeeds with 
probability one. Hence, the data compression limit, for the class of 
non-i.i.d. quantum information 
sources considered in this paper, is given by the von Neumann 
entropy rate $h$.

The following lemma shows that a compression scheme of rate $R < h$
is not reliable.

\begin{lemma}
\label{sufficient}
Let ${\cal S}_\Lambda$ be any set of eigenstates 
$\{|\psi_j^{\sigma, \Lambda}\rangle\}$ of 
$\rho^{\sigma, \Lambda}$ such that 
$$|{\cal S}_\Lambda| = 2^{[|\Lambda| R]},$$
where $R < h$ is fixed. Then for any $\epsilon > 0$ and sufficiently large
$\Lambda$
\be
\sum_{j\in {\cal S}_\Lambda }  \kappa_j^{\sigma, \Lambda} 
\le \epsilon. 
\label{bd4}
\ee
\end{lemma}
\begin{proof}
The LHS of \reff{bd4} gives the probability that an eigenstate
of $\rho^{\sigma, \Lambda}$ belongs to the set ${\cal S}_\Lambda$. 
We can write it as a sum of the probability that a state belonging
to ${\cal S}_\Lambda$ is $\delta$-typical and that it is atypical:
\be
\sum_{j\in {\cal S}_\Lambda }  \kappa_j^{\sigma, \Lambda} 
= \sum_{j\in {\cal S}_\Lambda}^{(\delta)} 
\kappa_j^{\sigma, \Lambda} 
+ {\sum_{j\in {\cal S}_\Lambda}^{(\delta)}}'
  \kappa_j^{\sigma, \Lambda} ;
\label{split}
\ee
here the second sum on the RHS of \reff{split} 
is over the atypical states in ${\cal S}_\Lambda$.
Choose $\delta >0$ such that $R < h-\delta$ and $0<\delta < \epsilon/2$.
In the limit ${\Lambda \nearrow\zed^d}$, the probability of atypical states 
is negligible. By \reff{kbound} the total probability of atypical states
can be made $< \,\epsilon$. 
There are
atmost $2^{[|\Lambda|R]}$ $\delta$-typical states in the set 
${\cal S}_\Lambda$, each with an eigenvalue  
$\le \,2^{-|\Lambda| (h - \delta)}$. Hence, the first term on RHS 
of \reff{split}
is bounded by 
$$
2^{-|\Lambda| (h - \delta)}\,2^{[|\Lambda|R]} 
\,\le\, 2^{-|\Lambda| \epsilon/2},
$$
which goes to zero in the limit $\Lambda \nearrow\zed^d$.

\end{proof}

We conclude this section with a
theorem giving giving the data compression limit and the 
limiting fidelity of the
compression scheme for general ({{not necessarily orthogonal}}) 
decompositions of $\rho^{\sigma, \Lambda}$.

Consider {\it any}
representation of the density matrix
$\rho^{\sigma, \Lambda}$:
$$\rho^{\sigma, \Lambda}=\sum\limits_i p_i^{\sigma, \Lambda}\,
|\phi_i^{\sigma, \Lambda}\rangle\langle \phi_i^{\sigma, \Lambda}|, $$
where $|\phi_i^{\sigma, \Lambda}\rangle \in {\cal H}^\sigma_\Lambda$ 
are arbitrary vectors of unit norm ({\em{not necessarily orthogonal or even
linearly independent}}), and $p_i^{\sigma, \Lambda}\ge 0$,
$\sum_i p_i^{\sigma, \Lambda} =1$. To apply the above data
compression scheme consider an orthogonal projection $\Pi:{\cal
H}_\Lambda^\sigma \to{\cal C}$, where ${\cal C}$
is a subspace of ${\cal H}_\Lambda^\sigma$ such that 
the vectors $|\Pi\phi_i^{\sigma ,\Lambda}
\rangle$ are either collinear or orthogonal for different
$i$ (some of them may be $0$).

If such a projection exists then, necessarily, the vectors spanning ${\cal
C}$
are eigenvectors of $\rho^{\sigma,\Lambda}$ and each non-zero vector
$|\Pi\phi_i^{\sigma,\Lambda}\rangle$ is collinear to one of these
eigenvectors. If we
take ${\cal C}$ to be the subspace ${\cal M}_\delta^{\sigma, \Lambda}$,
spanned by the $\delta$--typical states $|\psi_j^{\sigma, \Lambda}\rangle$
of $\rho^{\sigma,\Lambda}$, then to each non-zero
vector $|\Pi\phi_i^{\sigma ,\Lambda}\rangle$ we can assign
a quasiclassical state
$|{\underline x}\rangle$ associated with the
eigenvector $|\psi_j^{\sigma,\Lambda}\rangle$ collinear to
$|\Pi\phi_i^{\sigma ,\Lambda}\rangle$.
Here ${\underline x}$ is a binary string
of length $\leq [\log_2\,\left({\rm{dim}}\;{\cal C}\right)]+1$. In this
case, the compression scheme can be represented by
the two maps given below:
\bea
E &:& |\phi_i^{\sigma, \Lambda}\rangle \mapsto
|\psi_j^{\sigma, \Lambda}\rangle\quad {\hbox{where}}\,\,|\psi_j^{\sigma,
\Lambda}\rangle \in {\cal M}_\delta^{\sigma, \Lambda};
\label{onem}\\ {{C}} &:& |\psi_j^{\sigma, \Lambda}\rangle \mapsto
|{\underline x}^{(j)}\rangle\quad {\hbox{where}}\,\, {\underline x}^{(j)}
\in \{0,1\}^{r}
; \,\, r \le  [\log_2\,\left({\rm{dim}}\;{\cal C}\right)]+1.
\label{twom}
\eea

We use the symbols $E$ and $C$ for the maps (\ref{onem}) and (\ref{twom})
to denote encoding and compression. Note that
map $C$ is one--to--one. Hence, the quasiclassical state
$|{\underline x}^{(j)}\rangle$ can be decompressed unambiguously
to yield the $\delta$--typical state $|\psi_j^{\sigma, \Lambda}\rangle$.
However, map $E$ is not necessarily one--to--one. Consequently,
the original vector $|\phi_i^{\sigma, \Lambda}\rangle$ cannot be recovered
with certainty from the state $|\psi_j^{\sigma, \Lambda}\rangle$. Hence,
we
consider the following prescription for decoding the state
$|\psi_j^{\sigma, \Lambda}\rangle$ (denoted by the map $D$):
$$
D : |\psi_j^{\sigma, \Lambda}\rangle \mapsto |\phi_k^{\sigma,
\Lambda}\rangle,
$$
where $|\phi_k^{\sigma, \Lambda}\rangle$ satisfies the relation:
$$
\langle \phi_k^{\sigma, \Lambda}| \psi_j^{\sigma, \Lambda}\rangle
= \max_i\, \langle \phi_i^{\sigma, \Lambda}| \psi_j^{\sigma,
\Lambda}\rangle.
$$

The fidelity of such a coding--decoding scheme can be defined as:
\be
F_\Lambda := \sum_ip_i^{\sigma, \Lambda}\langle\phi_i^{\sigma, \Lambda}|
\Pi |\phi_i^{\sigma, \Lambda}\rangle.
\label{fidel}
\ee
The fidelity takes values between $0$ and $1$ and equals to unity only 
when all the states $|\Pi \phi_i^{\sigma, \Lambda}\rangle$ are 
correctly decoded. In the following theorem  we show that $F_\Lambda$ tends 
to unity as ${\Lambda \nearrow \zed^d}$.
\smallskip

\noindent
\begin{theorem}
\label{nonortho}

(i) Choose ${\cal C}$ to be the space of $\delta$-typical states of
$\rho^{\sigma, \Lambda}$:
$$
{\cal C} ={\cal M}_\delta^{\sigma, \Lambda} := {\rm{span}}\,\lbrace
|\psi_j^{\sigma, \Lambda}\rangle\, :\,\,
2^{-|\Lambda| (h - \delta)} \ge \kappa_j^{\sigma, \Lambda} \ge
2^{-|\Lambda| (h + \delta)} \rbrace,$$
where the $|\psi_i^{\sigma, \Lambda}\rangle$'s are orthonormal eigenstates
of $\rho^{\sigma, \Lambda}$ and $\kappa_j^{\sigma, \Lambda}$ are their
corresponding eigenvalues.
Let $\Pi$ be the orthoprojection ${\cal H}_\Lambda^\sigma\to
{\cal C}$.
The fidelity $F_\Lambda$ of the map $\Pi$, given by \reff{fidel}, 
approaches one:
$$\lim_{\Lambda \nearrow \zed^d}F_\Lambda := \lim_{\Lambda \nearrow \zed^d}
\sum\limits_i p_i^{\sigma, \Lambda}\langle \phi_i^{\sigma, \Lambda}
|\Pi  |\phi_i^{\sigma, \Lambda}\rangle
=1.$$
\smallskip

\noindent
(ii) If, for some subspace
$\displaystyle{{{\cal D}}\subseteq {\cal H}_\Lambda^\sigma}$,
the orthoprojection $\tilde\Pi$: ${\cal H}_\Lambda^\sigma\to
\displaystyle{{{\cal D}}}$ has
fidelity tending to one then
$$\lim_{\Lambda \nearrow \zed^d} \inf\frac{1}{|\Lambda|}
{\log}_2 \left( \dim {{{\cal D}}}\right)\ge h,$$
where $h$ is the von Neumann entropy rate. 
\end{theorem}
\proof 
To verify (i), write:
\bea
\sum_i p_i^{\sigma, \Lambda} \langle \phi_i^{\sigma, \Lambda}|
\Pi
|\phi_i^{\sigma, \Lambda}\rangle &=&
\sum_i p_i^{\sigma, \Lambda} \langle \phi_i^{\sigma, \Lambda} |
 \,\sum_j|\psi_j^{\sigma, \Lambda}\rangle \langle
\psi_j^{\sigma, \Lambda}|
\,{\mathbf 1}(\psi_j^{\sigma, \Lambda}\in{\cal C})
|\phi_i^{\sigma, \Lambda}\rangle \nonumber\\
&=&\sum_j\langle \psi_j^{\sigma, \Lambda}|
\sum_i p_i^{\sigma, \Lambda}|\phi_i^{\sigma, \Lambda}\rangle \langle
\phi_i^{\sigma, \Lambda}|{\mathbf 1}(\psi_j^{\sigma, \Lambda}\in
{\cal C})
|\psi_j^{\sigma, \Lambda}\rangle \nonumber\\
&=&
\sum_j \langle \psi_j^{\sigma, \Lambda}|\rho^{\sigma, \Lambda}|
\psi_j^{\sigma, \Lambda}\rangle {\mathbf 1}(\psi_j^{\sigma, \Lambda}\in
{\cal C})\nonumber\\
&=& \sum_j \kappa_j^{\sigma, \Lambda} {\mathbf 1} (|
\frac{-1}{|\Lambda|}\log\kappa_j^{\sigma, \Lambda}-h|\le
\delta)\to 1,
\eea
by Theorem \ref{mainresult}.
Property (ii) is checked in a similar fashion.

{\bf Remark.} The argument in the proof of Theorem
2 does not depend on the nature of the density matrix
$\rho_\Lambda^{\sigma ,\Lambda}$ or space ${\cal H}_\Lambda^\sigma$.
In a somewhat different context, a statement similar to
Theorem 2 was established in \cite{petz} (see also the
references therein).

\section{Proof of Theorem \ref{mainresult}}
\label{proofs}
In view of \reff{seven'}, eq. \reff{one5} is equivalent to
$$\lim_{\Lambda \nearrow\zed^d} \sum_j  \kappa_j^{\sigma, \Lambda}
\, {\mathbf{1}}\left(| c_0\,\beta  
\langle \psi_j^{\sigma, \Lambda} | \frac{H_\Lambda^\sigma}{|\Lambda|}| 
\psi_j^{\sigma, \Lambda} \rangle + \left( \frac{c_0}{|\Lambda|}\,
\log_2 \, \Xi^{\sigma, \Lambda} - h \right) | \le \delta \right) = 1.
$$
This fact, together with Proposition \ref{psresults} and eq. \reff{vnrate}
reduces the assertion of Theorem \ref{mainresult} to the following fact:
$\forall \, \delta > 0$
\be
 \lim_{\Lambda \nearrow\zed^d} \sum_j  \kappa_j^{\sigma, \Lambda}
\,{\mathbf{1}}\left(| \frac{1}{|\Lambda|}  
\langle \psi_j^{\sigma, \Lambda} | {H_\Lambda^\sigma}| 
\psi_j^{\sigma, \Lambda} \rangle -  g^{(\sigma)} | \ge 
\frac{c_0\,\delta}{\beta} \right) = 0,
\label{star}
\ee
where $g^{(\sigma)}$ is defined through \reff{gsig}.

Eq. \reff{star} is a Law of large numbers  for the random variables
$\langle \psi_j^{\sigma, \Lambda} | {H_\Lambda^\sigma}| 
\psi_j^{\sigma, \Lambda} \rangle$ (with respect to probability distributions
${\cal P}^{\sigma, \Lambda}$). In terms of characteristic functions,
\reff{star} is equivalent to the following lemma:
\smallskip

\noindent
\begin{lemma}\label{lln} 
 For $\beta$ large enough and $\lambda$ small enough, for any 
$t \in \bbbr$ the following limit exists:
\be
 \lim_{\Lambda \nearrow\zed^d} \varphi^{(\Lambda)} (t/{|\Lambda|}) = e^{itg^{(\sigma)}}
\label{rholim}
\ee
where $\varphi(\cdot)$ is defined through \reff{varphi} and 
$g^{(\sigma)}$ by \reff{gsig}.
\end{lemma}
\proof 

From \reff{fs.60} and \reff{varphi} we have that 
\be
\varphi^{\sigma, \Lambda} (t/|\Lambda|) =
\langle e^ {itH_\Lambda^\sigma/|\Lambda|} \rangle^\sigma_\Lambda =
\frac{\tr
\left(  e^{itH_\Lambda^\sigma/|\Lambda|}\, e^{-\beta H_\Lambda^\sigma}\right)}
{\Xi^{\sigma,\Lambda} }
\label{expr1}
\ee
Henceforth, we shall suppress the superscript $\sigma$
 from the notation $H^{\sigma}_{\Lambda}$ 
and $\Xi^{\sigma,\Lambda}$.

Expanding $e^{itH_\Lambda/|\Lambda|}$ on the RHS of \reff{expr1}
we obtain
\bea
\varphi^{(\Lambda)} (t/|\Lambda|) &=& \sum_{n=0}^\infty\frac{1}{n!}\,
\left(\frac{it}{|\Lambda|}\right)^n 
 {\tr\left(H_\Lambda^n e^{-\beta H_\Lambda}\right) \over
\Xi^\Lambda}
\nonumber\\
&=:& 1 +  \sum_{n=1}^\infty T_n
\label{expan}
\eea
Let us first estimate the term $T_1$.
\bea
T_1 &=& \frac{it}{|\Lambda|} \tr_{{\cal H}^\sigma_\Lambda}\left(H_\Lambda e^{-\beta
    H_\Lambda}
\right) /
\Xi^\Lambda \nonumber\\
&=& \frac{it}{|\Lambda|} \left( \sum_{X : \atop{ X \cap \Lambda \ne \emptyset}}
\tr_{{\cal H}^\sigma_\Lambda} \left(\Phi_X  e^{-\beta H_\Lambda}\right) \right)/\Xi^\Lambda
\nonumber\\
&=&\frac{it}{|\Lambda|} \sum_{j \in \Lambda}
\sum_{X \ni j\atop{X \cap \Lambda \ne \emptyset}} \frac{1}{|X|} \tr (\Phi_X 
e^{-\beta H_\Lambda}) / \Xi^\Lambda\nonumber\\
&=& \frac{it}{|\Lambda|} \sum_{j \in \Lambda} 
\tr\left( \Theta^\Lambda_{j}e^{-\beta
    H_\Lambda}\right)/\Xi^\Lambda
=\frac{it}{|\Lambda|} \sum_{j \in \Lambda}
\langle \Theta^\Lambda_{j} \rangle^{\sigma}_{\Lambda}
\eea
where
\be
\Theta^\Lambda_{j} := \sum_{X \ni j \atop{X \cap \Lambda \ne \emptyset}} 
\frac{1}{|X|} \Phi_X.
\label{theta}
\ee
Now
$$
\langle \Phi_X \rangle^\sigma_\Lambda := (\tr \,\Phi_X\, e^{-\beta
    H_\Lambda})/\Xi^\Lambda,$$
and 
\be
| \langle \Phi_X \rangle^\sigma_\Lambda| \le || \Phi_X||,
\label{bde1}
\ee
where $||\cdot ||$ denotes the Hilbert-Schmidt norm of the interaction
$\Phi_X$. Let
\be
c_0 := \max_{X:\atop{X \cap \Lambda \ne \emptyset}}|| \Phi_X|| .
\ee
Due to the finite range of the interaction, we have that
$$ \#\{ X \ni j| \Phi_X \ne 0, j \in \zed^d\} = \left(2^{2R}\right)^d,$$
for any site $j \in \zed^d$.  Hence,
\be
|\langle \Theta^\Lambda_{j}\rangle^\sigma_\Lambda|
\le \sum_{X \ni j\atop{X \subset \zed^d}} \frac{1}{|X|} |\langle\Phi_X
\rangle^\sigma_\Lambda|\le c_0 \, \left(2^{2R}\right)^d < \infty.
\label{thetabd}
\ee
It is known that for $\beta$ large enough and $\lambda$ small enough,
the following limit exists 
\be
\langle\Phi_X\rangle^\sigma:= \lim_{\Lambda\nearrow\zed^d}
\langle\Phi_X\rangle^\sigma_\Lambda
\ee
and defines the infinite volume Gibbs state \JMPcite{DFF, bku}.

Moreover,
\be
\langle \Theta_{0}\rangle^\sigma_{\zed^d}
:= \sum_{X \ni 0 \atop{X \subset \zed^d}} 
\frac{1}{|X|}\langle \Phi_X\rangle^\sigma
\equiv \sum_{X  \ni j \atop{ X\subset \zed^d}} 
\frac{1}{|X|} \langle\Phi_X\rangle^\sigma \quad \forall \,\, j \in \zed^d.
\ee
The last equality 
follows from the translational invariance of the interactions.

Further, by using methods of \JMPcite{DFF} it can be shown that
for $\beta$ large and $\lambda$ small enough, the following bound holds:
\be
|\langle \Phi_X \rangle^\sigma_\Lambda - \langle \Phi_X \rangle^\sigma|
\le ||\Phi_X||\, c(s(X))\, \Gamma\left(\dist(X, \partial \Lambda)\right);
\label{ninet}
\ee
here $\partial \Lambda$ denotes the boundary of the volume $\Lambda$,
$s(X)$ is the number of sites in the smallest connected
set of sites containing $X$, and the function $\Gamma (r)$, $r >0$, obeys
\be
| \Gamma(r)| \le \exp(-c_1 r),
\label{gammainf}
\ee
where $c_1>0$ is a constant depending on $\beta$ and $\lambda$. Note that 
$\Gamma$ does not depend on $\Lambda$. 

Using our assumptions on $\Phi$, one can prove the following Cesaro convergence:
\be
\lim_{\Lambda \nearrow\zed^d} T_1 \equiv 
\lim_{\Lambda \nearrow\zed^d}  \frac{it}{|\Lambda|} \sum_{j \in \Lambda} 
\langle \Theta^\Lambda_{j}\rangle^\sigma_\Lambda = it 
\langle \Theta_{0}\rangle^{\sigma}_{\zed^d}.
\label{lim1}
\ee
To prove \reff{lim1} consider $\Lambda$ to be a finite hypercubic volume 
$[-n,n]^d \cap \zed^d$ and define a subvolume $\widehat{\Lambda}$ as follows:
\be
\widehat{\Lambda} := \{ i \in \Lambda \,| \,\dist(i,j) \ge 
\log_e l(\Lambda)\, \forall
\,j \,\in \partial\Lambda\}.
\ee
Here $l(\Lambda)=2n +1$ is the linear size of the volume $\Lambda$. 
In the limit $\Lambda\nearrow\zed^d$, we have:
\be\label{lim2}\begin{array}{ll}
(a)\quad\frac{|\widehat{\Lambda}|}{|\Lambda|}
{{\longrightarrow}}\,\, 1 \quad \quad & (b) \quad
\frac{|\Lambda\setminus\widehat{\Lambda}|}{|\Lambda|}\equiv\frac{|\Lambda| 
- |\widehat{\Lambda}|}{|\Lambda|}\,\,
{{\longrightarrow}}\,\, 0 \\
\\
(c)\quad\frac{|\partial \widehat{\Lambda}|}{|\Lambda|}
{{\longrightarrow}}\, 0 \quad \quad & (d) \quad
\dist (\widehat{\Lambda}, \partial\Lambda) \,\,
{{\longrightarrow}}\,\, \infty.\end{array}
\ee
We can write
\bea
T_1 &=& \frac{it}{|\Lambda|} \sum_{j \in \Lambda}
\langle \Theta^\Lambda_{j} \rangle^\sigma_\Lambda
\nonumber\\
 &=&\frac{it}{|\Lambda|}\left[ \sum_{j \in \widehat{\Lambda}}
\langle \Theta^\Lambda_{j} \rangle^\sigma_\Lambda
+  \frac{it}{|\Lambda|} \sum_{j \in \Lambda\setminus \widehat{\Lambda}}
\langle \Theta^\Lambda_{j} \rangle^\sigma_\Lambda\right].
\label{t2}
\eea
Now
\be
\lim_{\Lambda\nearrow\zed^d}|\frac{it}{|\Lambda|} 
\sum_{j \in \Lambda\setminus \widehat{\Lambda}}
\langle \Theta^\Lambda_{j} \rangle^\sigma_\Lambda|  \le 
\lim_{\Lambda\nearrow\zed^d} {|t|}
\frac{|\Lambda\setminus \widehat{\Lambda}|}{|\Lambda|}
{\sup_{j \in \Lambda\setminus \widehat{\Lambda}}} 
|\langle \Theta^\Lambda_{j} \rangle^\sigma_\Lambda|.
\label{two4}
\ee
Hence, from \reff{thetabd} and (\ref{lim2}b)
\be
{\hbox{RHS of}} \quad \reff{two4} \le  
\lim_{\Lambda\nearrow\zed^d} c_0 |t| \left(2^{2R}\right)^d  
\frac{|\Lambda\setminus \widehat{\Lambda}|}{|\Lambda|}=0.
\ee

Consequently, in the infinite volume limit, the 
second term on the RHS of \reff{t2} goes to zero, thus allowing us 
to concentrate on the first term alone:
\bea
\lim_{\Lambda\nearrow\zed^d} T_1 &=& \lim_{\Lambda\nearrow\zed^d} 
\frac{it}{|\Lambda|}
\sum_{j \in \widehat{\Lambda}}
\langle \Theta^\Lambda_{j} \rangle^\sigma_\Lambda \nonumber\\
&=&\lim_{\Lambda\nearrow\zed^d} 
\left[\frac{it}{|\Lambda|} 
\sum_{j \in \widehat{\Lambda}}
\left[\langle \Theta^\Lambda_{j} \rangle^\sigma_\Lambda -
\langle \Theta_{0} \rangle^\sigma_{\zed^d}\right] +
\frac{it|\widehat{\Lambda}|}{|\Lambda|}
\langle \Theta_{0} \rangle^\sigma_{\zed^d}\right]  \nonumber\\
&=& it \langle \Theta_{0} \rangle^\sigma + A,
\label{twen6}
\eea
where
$$A := \lim_{\Lambda\nearrow\zed^d} 
\frac{it}{|\Lambda|} 
\sum_{j \in \widehat{\Lambda}}
\left[\langle \Theta^\Lambda_{j} \rangle^\sigma_\Lambda -
\langle \Theta_{0} \rangle^\sigma_{\zed^d}\right].
$$
The last line of \reff{twen6} follows from (\ref{lim2}a).
We shall prove that $A=0$. Write 
\bea
A &=&\lim_{\Lambda\nearrow\zed^d}  \left[ \frac{it}{|\Lambda|} 
\sum_{j \in \widehat{\Lambda}}
\sum_{X \ni j\atop{X \subset \Lambda \cup \Lambda^\partial}}
\frac{1}{|X|} \left(
\langle \Phi_X \rangle^\sigma_\Lambda - \langle \Phi_X \rangle^\sigma\right)
-  \frac{it}{|\Lambda|}\sum_{j \in \widehat{\Lambda}}
\sum_{X \ni j\atop{X \not\subset \Lambda\cup \Lambda^\partial }}\frac{1}{|X|} 
\langle \Phi_X \rangle^\sigma\right]\nonumber\\
&:=& A_1 + A_2.
\label{four6}
\eea
Recall that the interaction governing the system is of 
a finite range $R$. Define:
$$
{\widehat{\Lambda}}^{(R)}_j := \{ i \in \Lambda | 
\dist(i,j) \le R\}, \quad \,\,j \in \widehat{\Lambda}.
$$
Then we have
\be
|A_1| \le \lim_{\Lambda\nearrow\zed^d} \left[ \frac{|t|}{|\Lambda|}
\sum_{j \in \widehat{\Lambda}} \sum_{X \subset {\widehat{\Lambda}}^{(R)}_j}
\frac{1}{|X|} 
|\langle \Phi_X \rangle^\sigma_\Lambda - \langle \Phi_X \rangle^\sigma|\right].
\label{twen9}
\ee
Using \reff{ninet} we obtain
$$
|A_1| \le \lim_{\Lambda\nearrow\zed^d}\frac{|t|}{|\Lambda|}
\sum_{j \in \widehat{\Lambda}} \sum_{X
\subset {\widehat{\Lambda}}^{(R)}_j}
\frac{1}{|X|} ||\Phi_X||\, c(s(X))\, \Gamma(\dist(X, \partial\Lambda)).
$$
Set: 
$$
c_2 := {\sup_{{X \subset {\widehat{\Lambda}}^{(R)}_j}}}\, c(s(X)).
$$
We have that
$$
\#\{ X| X \subset {\widehat{\Lambda}}^{(R)}_j\} = \left(2^{2R}\right)^d,
$$
and for $ X \subset {\widehat{\Lambda}}^{(R)}_j$,
$$
\Gamma(\dist(X, \partial\Lambda)) \le \Gamma(\dist(\widehat{\Lambda}, 
\partial\Lambda) - R).
$$
Hence,
\be
|A_1| \le
\lim_{\Lambda\nearrow\zed^d}|t|\,\frac{|\widehat{\Lambda}|}{|\Lambda|}
\,c_0\,c_2 \,\left(2^{2R}\right)^d \, 
\Gamma(\dist(\widehat{\Lambda}, 
\partial\Lambda) - R)=0.
\label{four8}
\ee
by (\ref{lim2}d) and \reff{gammainf}.

The second term on the RHS of \reff{four6} is bounded as follows:
\bea
|A_2| &\le&  \lim_{\Lambda\nearrow\zed^d}\frac{|t|}{|\Lambda|}
\sum_{j \in \widehat{\Lambda}}\sum_{X \ni j\atop{X \not\subset \Lambda\cup \Lambda^\partial}}\frac{1}{|X|}|\langle \Phi_X \rangle^\sigma|\nonumber\\
 &\le&  \lim_{\Lambda\nearrow\zed^d}\frac{|t|}{|\Lambda|}
\sum_{j \in \widehat{\Lambda}}\sum_{X \ni j\atop{X \not\subset 
\widehat{\Lambda}}}\frac{1}{|X|}|\langle \Phi_X \rangle^\sigma|,
\eea
since $\widehat{\Lambda} \subset \Lambda$. Now
$$
\#\{ X| X \ni j, X \not\subset {\widehat{\Lambda}}, j 
\in \widehat{\Lambda}, \Phi_X \ne 0\} 
= \left(2^{2R}\right)^d
$$
and
$$\#\{j\in \widehat{\Lambda}|\, \exists \, X \ni j, \,
{\hbox{such that}}  \, X \not\subset {\widehat{\Lambda}}, \Phi_X \ne 0\} = 
| {\hbox{Int}}^{(R)} ( \widehat{\Lambda}) |,
$$
where 
$$
{\hbox{Int}}^{(R)} ( \widehat{\Lambda}) := \{ i \in  \widehat{\Lambda}
| \dist (i, \partial\widehat{\Lambda}) < R\}
$$ is the $R$-interior of the volume $ \widehat{\Lambda}$.
We have
$$
|A_2| \le \lim_{\Lambda\nearrow\zed^d}\frac{|t|}{|\Lambda|}
\,\left(2^{2R}\right)^d\, c_0 \,| {\hbox{Int}}^{(R)} ( \widehat{\Lambda}) |.
$$
However,
$$
| {\hbox{Int}}^{(R)} ( \widehat{\Lambda}) | \le R \, |\partial 
\widehat{\Lambda}|.
$$
Hence,
\bea
|A_2| &\le& \lim_{\Lambda\nearrow\zed^d}|t|\,\frac{|\partial \widehat\Lambda|}
{|\Lambda|} \,R\,c_0\,\left(2^{2R}\right)^d \nonumber\\
&=& 0,
\label{three4}
\eea
by (\ref{lim2}c). From \reff{twen6}, \reff{four6}, \reff{four8} and \reff{three4} we 
readily get \reff{lim1}.

This argument admits a generalisation for the 
$n^{th}$ term in the expansion on the RHS of \reff{expan}. We have:
\bea
T_n &:=& \frac{1}{n!} \left(\frac{it}{|\Lambda|}\right)^n 
 \tr\left(H_\Lambda^n e^{-\beta H_\Lambda}\right) /
\Xi^\Lambda \nonumber\\
&=& \frac{1}{n!} \left(\frac{it}{|\Lambda|}\right)^n \sum_{X_1, \ldots,X_n
\subset\atop{X_i \cap \Lambda \ne \emptyset}} \tr \left(\Phi_{X_1}\ldots \Phi_{X_n}e^{-\beta H_\Lambda}
\right) / \Xi^\Lambda \nonumber\\
&=& \frac{1}{n!} \left(\frac{it}{|\Lambda|}\right)^n
\sum_{j_1\ldots j_n \in \Lambda} \sum_{X_1 \ni j_1\atop{X_1 \cap
    \Lambda
\ne \emptyset}}
\cdots  \sum_{X_n \ni j_n\atop{X_n \cap \Lambda \ne \emptyset}}\frac{1}{|X_1|}\ldots
\frac{1}{|X_n|} \langle \Phi_{X_1}\ldots \Phi_{X_n}\rangle^\sigma_\Lambda
\nonumber\\ 
&=& \frac{1}{n!} \left(\frac{it}{|\Lambda|}\right)^n
\sum_{j_1\ldots j_n \in \Lambda}\langle \Theta^{\Lambda}_{j_1}\ldots 
 \Theta^{\Lambda}_{j_n}\rangle^\sigma_\Lambda.
\label{three6}
\eea
We prove below that for each $n\ge 2$,
\be
\lim_{\Lambda\nearrow\zed^d} T_n = \frac{(it)^n}{n!} \left(\langle \Theta_0
\rangle^\sigma_{\zed^d}\right)^n.
\label{three3}
\ee

Define volumes $\Lambda^{(n)}$ and ${\widehat{\Lambda}}^{(n)}$:
\bea
\Lambda^{(n)} &=& \{(i_1,\ldots, i_n) | \, i_k \in \Lambda \,\forall\,
\, 1 \le k \le n\},\nonumber\\
{\widehat{\Lambda}}^{(n)} &=& \{(j_1,\ldots, j_n) | \, (j_1, \ldots j_n )
\in \Lambda^{(n)}, L_n \ge \ell_\Lambda^{(n)}\},
\eea
where
$$
L_n \equiv L(j_1,\ldots, j_n) := \min\left\{{\min_{1\le k<l\le n}} \, 
\left[\dist(j_k, j_l)\right]\, , \, 
{\min_{1\le k\le n}} \, \left[\dist(j_k, \partial\Lambda )\right] \right\},
$$
and $\partial\Lambda$ is the boundary of the volume 
$\Lambda$. The quantity $\ell_\Lambda^{(n)} $ is chosen so that
\be
\lim_{\Lambda\nearrow\zed^d}\,\ell_\Lambda^{(n)} = \infty
\label{rinf}
\ee
and 
\be \label{lim3}
(a)\,\,\frac{|{\widehat{\Lambda}}^{(n)}|}{|\Lambda^{(n)}|}
{{\longrightarrow}}\,\, 1, \quad \quad (b) \,\,
\frac{|\Lambda^{(n)}  \setminus {\widehat{\Lambda}}^{(n)}|}{|\Lambda^{(n)}|}\,\,
{{\longrightarrow}}\,\, 0, \quad \quad
(c)\,\,\frac{|\partial {\widehat{\Lambda}}^{(n)}|}{|\Lambda^{(n)}|}
{{\longrightarrow}}\,\, 0. \quad \quad 
\ee
\smallskip

\noindent
[Note that $|\Lambda^{(n)}| = |\Lambda|^n$.] Writing 
$\jj = (j_1, \ldots , j_n)$, we prove \reff{three3} as follows:
\bea
T_n &:=& \frac{1}{n!} \left(\frac{it}{|\Lambda|}\right)^n 
\left[ \sum_{\jj \in {\widehat{\Lambda}}^{(n)} }
\langle \Theta^{\Lambda}_{j_1}\ldots 
 \Theta^{\Lambda}_{j_n}\rangle^\sigma_\Lambda
+  \sum_{\jj \in {\Lambda^{(n)} \setminus \widehat{\Lambda}}^{(n)}} 
\langle \Theta^{\Lambda}_{j_1}\ldots 
 \Theta^{\Lambda}_{j_n}\rangle^\sigma_\Lambda \right] \nonumber\\
&:=& T_n(1) + T_n (2).
\label{four0}
\eea
Now,
\bea
|\langle \Theta^{\Lambda}_{j_1}\ldots 
 \Theta^{\Lambda}_{j_n}\rangle^\sigma_\Lambda|&\le&
 \sum_{X_1 \ni j_1\atop{X_1 \cap \Lambda \ne \emptyset}}
\cdots  \sum_{X_n \ni j_n\atop{X_n \cap \Lambda \ne \emptyset}}
\frac{1}{|X_1|}\ldots
\frac{1}{|X_n|}| \langle \Phi_{X_1}\ldots \Phi_{X_n}\rangle^\sigma_\Lambda|
\nonumber\\
&\le&\left[\left(2^{2R}\right)^d\right]^n\, || \Phi_{X_1}\ldots \Phi_{X_n}||\nonumber\\
&\le& c_0^n \, \left(2^{2R}\right)^{nd}.
\eea
Hence, 
\bea
\lim_{\Lambda\nearrow\zed^d} T_n(2) &\le& \lim_{\Lambda\nearrow\zed^d}
\frac{|\Lambda^{(n)}  \setminus {\widehat{\Lambda}}^{(n)}|}{|\Lambda^{(n)}|} 
\frac{|t|^n}{n!}  c_0^n \, \left(2^{2R}\right)^{nd} \nonumber\\
&=& 0,
\eea
by (\ref{lim3}c). Consequently, in the infinite volume limit, the only
non-zero contribution to $T_n$ arises from the term
$T_n(1)$ on the RHS of \reff{four0}. This term can 
in turn can be written as follows:
\bea
\lim_{\Lambda\nearrow\zed^d} T_n(1) &\le& \frac{1}{n!}\,
\left(\frac{it}{|\Lambda|}\right)^n\left[
\sum_{j \in {\widehat{\Lambda}}^{(n)}}
\left[\langle \Theta^\Lambda_{j_1}\ldots\Theta^\Lambda_{j_n}  
\rangle^\sigma_\Lambda -
(\langle \Theta_{0} \rangle^\sigma_{\zed^d})^n\right]
 +
\left(\langle \Theta_{0} \rangle^\sigma_{\zed^d}\right)^n \right] \nonumber\\
&=& \frac{(it)^n}{n!}\, 
\left(\langle \Theta_{0} \rangle^\sigma_{\zed^d}\right)^n \, + \, B,
\label{six0}
\eea
where
$$B := \lim_{\Lambda\nearrow\zed^d} T_n(2) \le \frac{1}{n!}\,
\left(\frac{it}{|\Lambda|}\right)^n
\sum_{j \in {\widehat{\Lambda}}^{(n)}}
\left[\langle \Theta^\Lambda_{j_1}\ldots\Theta^\Lambda_{j_n}  
\rangle^\sigma_\Lambda -
(\langle \Theta_{0} \rangle^\sigma_{\zed^d})^n\right].
$$
The first term on the RHS of the last line of \reff{six0} follows 
from (\ref{lim3}a). We prove below that
$B=0$. We can write $B$ as follows:
\bea
B &:=& \lim_{\Lambda\nearrow\zed^d} \,\frac{1}{n!}\,
\left(\frac{it}{|\Lambda|}\right)^n\sum_{j \in {\widehat{\Lambda}}^{(n)}}
 \Biggl[\sum_{X_1 \ni j_1\atop{X_1 \subset \Lambda \cup \Lambda^\partial}}
\cdots  \sum_{X_n \ni j_n\atop{X_n \subset \Lambda\cup\Lambda^\partial}}
\frac{1}{|X_1|}\ldots \frac{1}{|X_n|} \nonumber\\
& & \quad\times \Bigl(\langle \Phi_{X_1}\ldots \Phi_{X_n}\rangle^\sigma_\Lambda
 - \langle \Phi_{X_1}\ldots \Phi_{X_n}\rangle^\sigma \Bigr)
 + \sum_{X_1 \ni j_1} \cdots  \sum_{X_n \ni j_n} 
\frac{1}{|X_1|}\ldots
\frac{1}{|X_n|} \nonumber\\
& & \quad \times\langle \Phi_{X_1}\ldots \Phi_{X_n}\rangle^\sigma
{\mathbf{1}}(X_i \not\subset \Lambda \cup \Lambda^\partial \, 
{\hbox{for some}} \,
         1 \le i \le n)\Biggr]
\nonumber\\
&:=& B_1 + B_2.
\label{six1}
\eea
\smallskip

By using methods of \JMPcite{DFF} it can be shown that
for $\beta$ large and $\lambda$ small enough, the following bound holds:
$$
\left|\langle \prod_{i=1}^n \Phi_{X_i}\rangle^\sigma_\Lambda  - \langle
\prod_{i=1}^{n}
\Phi_{X_i}\rangle^\sigma\right| \le \left(\prod_{i=1}^{n} ||\Phi_{X_i}|| 
\right)\,
c(s(X_1, \ldots , X_n)\, \Gamma(\Delta_n),
$$
where
$$
\Delta_n \equiv \Delta_n(X_1, \ldots, X_n) = \min\{\dist(X_i, X_j) | 1
\le i <j \le n\},
$$ 
and $\Gamma(r)$ is a monotonically decreasing function of $r$, 
satisfying the bound \reff{gammainf}.
Further, recall that 
$||\Phi_{X_i}|| \le c_0$ and let $c_3 := {\sup_{ X_1\ldots X_n \subset \zed^d}}
\,c(s(X_1, \ldots, X_n)).$ 
For $X_i \subset \Lambda$ and $X_i \ni j_i$ for $1\le i \le n$,
$$
{\min_{1\le i <j \le n}} \, \{ \dist (X_i, X_j)\} \ge 
(\ell_\Lambda^{(n)} - 2R).$$
Hence, 
$$ \Gamma(\Delta_n) \le\Gamma (\ell_\Lambda^{(n)} - 2R),$$
and
\be
|B_1| \le \lim_{\Lambda\nearrow\zed^d} 
\,\frac{|t|}{n!} \, \frac{|{\widehat{\Lambda}}^{(n)}|}{|\Lambda^{(n)}|} 
c_3\,(c_0)^n 
\left(2^{2R}\right)^{nd}\,\Gamma (\ell_\Lambda^{(n)}  - 2R)
=0,
\label{six2}
\ee
by \reff{rinf} and \reff{gammainf}. Moreover, 
\bea
|B_2| &\le& \lim_{\Lambda\nearrow\zed^d}
\frac{|t|^n}{|\Lambda^{(n)}|}\,\frac{1}{n!} \,
\sum_{\jj \in {\widehat{\Lambda}}^{(n)} }
 \sum_{X_1 \ni j_1}\cdots  \sum_{X_n \ni j_n}
||\Phi_{X_1} \ldots  \Phi_{X_n}|| \, {\mathbf{1}} (X_i \not\subset \Lambda
\cup \Lambda^\partial \, {\hbox{ for some}} \, 1\le i \le n) \nonumber\\
&\le& \lim_{\Lambda\nearrow\zed^d}
\frac{|t|^n}{|\Lambda^{(n)}|}\,\frac{1}{n!} \,
\sum_{\jj \in {\widehat{\Lambda}}^{(n)} }
 \sum_{X_1 \ni j_1}\cdots  \sum_{X_n \ni j_n}
||\Phi_{X_1} \ldots  \Phi_{X_n}|| \, {\mathbf{1}} (X_i \not\subset
{\widehat{\Lambda}} 
\, {\hbox{ for some}} \, 1\le i \le n) \nonumber\\
& \le& \lim_{\Lambda\nearrow\zed^d}
\frac{|t|^n}{|\Lambda^{(n)}|} c_0^n \,\left(2^{2R}\right)^{nd} \,
| {\hbox{Int}}^{(R)} ( {\widehat{\Lambda}}^{(n)}) |.
\eea
In fact,
$$ \#\{ \jj \in {\widehat{\Lambda}}^{(n)} | \,
\exists \, X_1 \ni j_1,\ldots  {X_n \ni j_n}\,\,
{\hbox{such that}}\,  \, X_i \not\subset {\widehat{\Lambda}}\,
\,{\hbox{for some}}\,\, 1 \le i \le n\} = 
| {\hbox{Int}}^{(R)} ( {\widehat{\Lambda}}^{(n)}) |.
$$
Here, as before
$$
{\hbox{Int}}^{(R)} ( {\widehat{\Lambda}}^{(n)})
= \{ \jj \in {\widehat{\Lambda}}^{(n)} | \, \dist( \jj, \partial 
{\widehat{\Lambda}}^{(n)}) < R\}.
$$

Hence,  $ | {\hbox{Int}}^{(R)} ( {\widehat{\Lambda}}^{(n)}) |
\le R \, |\partial  {\widehat{\Lambda}}^{(n)} |  $
and 
\be
|B_2|\le   \lim_{\Lambda\nearrow\zed^d}\frac{|\partial  
{\widehat{\Lambda}}^{(n)} |}{|\Lambda|^n} \, 
\frac{|t|^n}{n!}\, c_0^n \,R\, \left(2^{2R}\right)^{nd}
= 0,
\label{six4}
\ee
by (\ref{lim3}b). From \reff{six1}, \reff{six2} and \reff{six4} 
it follows that $B=0$. Hence, from \reff{six0} and \reff{lim1} 
one obtains
\be
 \lim_{\Lambda\nearrow\zed^d} T_n = \frac{(it)^n}{n!} \, 
\left(\langle \Theta_0\rangle^\sigma_{\zed^d}\right)^n \quad \,\,
\forall n \ge 1.
\label{tlim}
\ee
From \reff{expan} we now see, in view of Lebesgue's dominated convergence
theorem, that 
\bea
 \lim_{\Lambda \nearrow\zed^d} \varphi^{(\Lambda)} (t/{|\Lambda|}) = 
1 + \sum_{n=1}^\infty \frac{(it)^n}{n!} \,
\left(\langle \Theta_0\rangle^\sigma_{\zed^d}\right)^n .
\label{six6}
\eea
The limiting energy density per lattice site, $g^{(\sigma)}$,
 defined through \reff{gsig}, can be written as
\be
g^{(\sigma)} :=  \lim_{\Lambda \nearrow\zed^d} \langle
H_\Lambda/|\Lambda|\rangle^\sigma_\Lambda \,=\,  \lim_{\Lambda \nearrow\zed^d} \sum_{X:\atop{X\cap \Lambda
\ne \emptyset}} 
\frac{\langle \Phi_X \rangle^\sigma_\Lambda}{|\Lambda|} 
\,=\, \lim_{\Lambda \nearrow\zed^d}\frac{1}{|\Lambda|} 
\sum_{j \in \Lambda} \sum_{X \ni j \atop{X \cap\Lambda
\ne \emptyset}} 
\frac{\langle \Phi_X \rangle^\sigma_\Lambda}{|X|}.
\ee
Since the interaction $\Phi = \{\Phi_X\}$ is assumed to be translationally
invariant we can write
\be
g^{(\sigma)} = \lim_{\Lambda \nearrow\zed^d}\frac{1}{|\Lambda|} 
\sum_{j \in \Lambda} \sum_{X \ni 0\atop{X \cap \Lambda
\ne \emptyset}} 
\frac{\langle \Phi_X \rangle^\sigma_\Lambda}{|X|}\,
=\,  \lim_{\Lambda \nearrow\zed^d}
\sum_{X \ni 0\atop{ X \cap \Lambda \ne \emptyset}} 
\frac{\langle \Phi_X \rangle^\sigma_\Lambda}{|X|}\,
=\, \sum_{X \ni 0 \atop{X \subset \zed^d}} 
\frac{\langle \Phi_X \rangle^\sigma}{|X|}
\,=\, \langle \Theta_0 \rangle^\sigma_{\zed^d}.
\ee
Hence, \reff{six6} can be written as 
$$
 \lim_{\Lambda \nearrow\zed^d} \varphi^{(\Lambda)} (t/{|\Lambda|})
 = e^{itg^{(\sigma)}},
$$
which proves Lemma \ref{lln}.
\section*{Acknowledgements}
This work has been done in association with the 
Cambridge--Massachusetts--Institute (CMI).
YMS thanks I.H.E.S., Bures-sur-Yvette, France,
Laboratoire de Probabilit\'e, Universit\'e
Paris-6 (P. et M. Curie)
and DIAS, Dublin, for support and hospitality during
visits in 2001 and 2002. We would like thank A. Kaltchenko
for pointing out a mistake in the original manuscript.                       
\bigskip

\noindent
{
\def\thebibliography#1{\noindent
\section*{References}\par \list
{${\hbox{\arabic{enumi}.}}$}{\settowidth\labelwidth{[#1]} 
\leftmargin\labelwidth
\advance\leftmargin\labelsep \itemsep=0pt \usecounter{enumi}}
\def\newblock{\hskip .11em plus .33em minus -.07em} \sloppy
\sfcode`\.=1000\relax}

\end{document}